# Moiré exciton dynamics and moiré exciton-phonon interaction in a WSe$_2$/MoSe$_2$ heterobilayer

Keisuke Shinokita[1,†], Yuhei Miyauchi[1], Kenji Watanabe[2],

Takashi Taniguchi[3], and Kazunari Matsuda[1,*]

[1]*Institute of Advanced Energy, Kyoto University, Uji, Kyoto 611-0011, Japan*

[2]*Research Center for Functional Materials, National Institute for Materials Science, 1-1 Namiki, Tsukuba, Ibaraki 305-0044, Japan*

[3] *International Center for Materials Nanoarchitectonics, National Institute for Materials Science, 1-1 Namiki, Tsukuba, Ibaraki 305-0044, Japan*

## Abstract

Moiré patterns with angular mismatch in van der Waals heterostructures composed of atomically thin semiconducting materials are a fascinating platform to engineer the optically generated excitonic properties towards novel quantum phenomena. The moiré pattern as a periodic trap potential can give rise to spatially ordered zero-dimensional (0D) exciton ensembles, which offers the possibility for dense coherent quantum emitters and quantum simulation of many-body physics. The intriguing moiré exciton properties are affected by their dynamics and exciton-phonon interaction. However, the moiré exciton dynamics and the interaction between the moiré exciton and phonon are still elusive. Here, we report the moiré exciton and phonon interaction in a twisted WSe$_2$/MoSe$_2$ heterobilayer based on near-resonant photoluminescence excitation spectroscopy. We observed the selective excitation of the ground state of the moiré exciton at phonon resonance. The otherwise negligible small absorption below the continuum state is a hallmark of the density of states of a 0D-like system. In addition, the excitation power dependence of the PL spectra reveals the dynamics of moiré exciton ensembles between different potential minima with discrete energy levels via the resonant phonon scattering process. The results presented here of the moiré exciton dynamics under suppressed phonon interaction could pave a new way for the exploration of novel quantum phenomena of the moiré exciton towards potential applications in quantum optics.

†shinokita.keisuke.4r@kyoto-u.ac.jp

*matsuda@iae.kyoto-u.ac.jp

The interference of two similar patterns is a universal concept in physics that plays a pivotal role in modern science and technology such as in gravitational wave detection[1], optical frequency combs[2], superconducting quantum interference devices (SQUIDs)[3], and cold atoms in optical lattices[4]. The moiré patterns of van der Waals heterostructures arising from interference of angular- or lattice-mismatched atomically thin materials with honeycomb structures, such as graphene and semiconducting transition metal dichalcogenides (TMDs), have attracted increasing attention because of the potential for engineering a range of emergent quantum phenomena. Examples include superconductivity[5,6], ferromagnetism near ¾ filling[7], and correlated insulator phases[8] in twisted bilayer graphene. In a two-dimensional (2D) semiconducting TMD heterostructure, the stacking of two different monolayer TMDs usually results in staggered type II band alignment, which causes separation of electrons and holes in different layers, or interlayer excitons (Coulomb-bound electron-hole pairs)[9–15]. The nature of the interlayer excitons is modulated by the moiré pattern because of the spatially varying atomic registry[16–25]. The moiré pattern works as a periodic trap potential to confine the interlayer exciton in zero dimensions (0D) (moiré exciton, Fig. 1a) and spatially organize the moiré-trapped excitons, which results in an array of quantum-dot-like 0D systems composed of a moiré exciton ensemble. In addition, the moiré period and interaction between the moiré excitons can be tailored by the stacking angle. Therefore, moiré exciton ensembles in periodic moiré potentials have potential for dense coherent quantum emitters and quantum simulation of many-body physics[25,26], which could result in a number of applications in quantum optics, including quantum dot lasers, entangled photon lasers[27], and Dicke superradiance[28].

The electronic, optical, and transport properties of solids are frequently dominated by electron-phonon or exciton-phonon interactions. Extensive works on exciton and phonon interactions in 2D materials and their van der Waals heterostructures have been completed[29–35]. For instance, emergent interlayer exciton-phonon coupling was observed in a $WSe_2$/$h$-BN heterostructure system[32], which provides important information for the generation and control of intriguing physical properties of 2D materials. The exciton-phonon interaction of the heterostructure can also be modified by the periodic moiré potential, which would pave a new way for control of diverse fascinating physical behaviours of 0D-like moiré excitons towards coherent quantum emitters and quantum simulation of many-body physics. To date, the signatures of moiré excitons have been optically studied by absorption and photoluminescence (PL) measurements[17–20], where the moiré exciton was confirmed by the appearance of sharp peaks in low-temperature PL spectra under low excitation power conditions, reflecting the trapping of excitons in the moiré potential. However, the dynamics of the moiré exciton and interaction between the moiré exciton and phonon have yet to be studied experimentally. To explore novel quantum phenomena in moiré superlattices, it is important to understand the moiré exciton dynamics and moiré exciton-phonon interaction, which play a dominant role in the intriguing properties of moiré exciton ensembles and quantum applications.

Here, we study the moiré exciton and phonon interaction in a twisted $WSe_2$/$MoSe_2$ heterobilayer based on near-resonant photoluminescence excitation (PLE) spectroscopy. The experimentally observed PL spectrum strongly depending on the excitation energy shows highly selective excitation of the ground state of the moiré exciton ($n$=1) at phonon resonances. On the other hand, the negligibly small off-resonant PLE signal in the interlayer region and the absence of excited states (*n*=2, 3, …) of the moiré exciton suggest *δ*-function discrete energy levels (Fig. 1b), which are a hallmark of the density of states of a 0D-like system for the interlayer moiré exciton. In addition, the excitation power dependence of the PL spectra reveals the moiré exciton dynamics between different potential minima with discrete energy levels via the resonant phonon scattering process.

The twisted WSe$_2$/MoSe$_2$ heterobilayer studied here was prepared using the polymer stamp dry-transfer technique. The mechanically exfoliated monolayer (1L-) WSe$_2$, 1L-MoSe$_2$, and *h*-BN on a SiO$_2$/Si substrate were picked up and stacked using a poly(vinyl chloride) (PVC) sheet on a polydimethylsiloxane (PDMS) lens[36,37]. The relatively large stacking angle of 1L-MoSe$_2$ and 1L-WSe$_2$ in the twisted WSe$_2$/MoSe$_2$ heterobilayer was evaluated to be 10±1° by the polarization dependence of the second harmonic generation (see Supplementary Information). The moiré period with a twist angle of *θ*=10° was estimated to be $a_{moiré} = a/2\sin(\theta/2)$=2 nm, where *a*=0.328 nm is the lattice constant. The twisted WSe$_2$/MoSe$_2$ heterobilayer was encapsulated by top and bottom *h*-BN layers. The bottom and top *h*-BN thicknesses were ~20 and ~10 nm, respectively. Figure 1c shows an optical image of the heterobilayer used in this study. The red and blue surrounding areas are regions of 1L-WSe$_2$ and 1L-MoSe$_2$, respectively. The overlapping area of 1L-WSe$_2$ and 1L-MoSe$_2$ corresponds to the heterobilayer region.

A tuneable continuous-wave (cw) Ti:sapphire laser (photon energy from 1.35 to 1.67 eV, bandwidth of 0.3 meV) was used as the excitation source for PLE measurements. The laser was focused using a 100× objective lens, and the spot size on the sample was ~2 μm. The light emission from the twisted $MoSe_2/WSe_2$ was collected using the same objective lens and detected with a cooled charge-coupled device (CCD) through a spectrometer with a spectral resolution of 0.6 meV. All measurements were performed at 10 K with a cryogen-free cryostat.

Figure 1d shows the PL spectrum (red curve) under far-off-resonance excitation at a photon energy of 2.33 eV with a Nd:YAG laser at 10 K. The PL spectrum shows strong peaks at approximately 1.34 and 1.6 eV, which are assigned to the emissions from interlayer and intralayer excitons, respectively[9,38]. Two peaks are observed in the intralayer exciton region of $MoSe_2$ originating from neutral and negatively charged excitons ($X^0$ and $X^-$, respectively) at 1.644 and 1.617 eV, respectively. Figure 1e shows the PL spectra of the twisted $MoSe_2/WSe_2$ heterobilayer at different excitation densities. While at a high excitation intensity, a single smooth peak is dominant, at a low excitation intensity, a more complicated structure with additional peaks, including at 1.32 and 1.33 eV, appears. This behaviour is consistent with previously reported results and reflects the response of the interlayer exciton trapped with the moiré superlattice[17–20].

The PLE spectrum probed around 1.34 eV of the interlayer exciton emission signal is shown in Fig. 1d (blue curve). The PLE signal from the interlayer exciton was enhanced when the excitation energy was resonant with the intralayer exciton of monolayer $MoSe_2$ around 1.6 eV, where the interlayer exciton was formed via energy relaxation from the intralayer exciton created in the $MoSe_2$ layer[39]. In contrast, once the photoexcitation energy is below the intralayer exciton energy of 1.6 eV, the PLE signal decreases drastically: The PLE signal in the interlayer exciton region below 1.4 eV is 3-4 orders of magnitude smaller than that in the intralayer exciton region above 1.6 eV. The drastic decrease in the PLE signal reflects the smaller oscillator strength of the interlayer exciton with a reduced dipole moment because of the spatially separated electron and hole in different layers[39–41]. The PLE signal at the intralayer exciton energy of approximately 1.6 eV has a tail on the lower energy side. A comparison between the PLE and PL signals reveals that the PLE tail contains information on the density of states of the moiré exciton, which will be discussed in detail later. The PL linewidth under the resonant condition (Fig. 2a) is broader than the excitation laser linewidth of 0.3 meV and spectral resolution of 0.6 meV, which suggests that the observed signals originate not from the resonant Raman signal but from the resonant PL signal.

To obtain more detailed information on the energy levels of the interlayer moiré exciton, we performed PLE spectroscopy under near-resonant photoexcitation conditions. Figure 2a shows the PL spectra with different excitation energies. The PL spectral shape strongly depends on the excitation energy under the near-resonant excitation conditions: The higher (lower) energy interlayer moiré exciton emission around 1.34 (1.33) eV was enhanced under an excitation energy of approximately 1.37 (1.36) eV, suggesting highly selective excitation of a specific moiré exciton under resonance excitation conditions. The resonance behaviour is clearly shown in the 2D colour map of excitonic emissions as a function of laser excitation energy in Fig. 2b. The PLE signal shows pronounced resonance features along the tilted dashed lines with excess energies of 24 and 48 meV. To quantitatively discuss the resonance features, we fitted the PL spectrum in Fig. 2a with multiple Lorentz functions with a linewidth of 3 meV (black lines) and extracted the PL intensity of each Lorentz function (see Supplementary Information). Figure 2c shows the fitting amplitudes of Lorentz functions at centre energies of 1.334, 1.338, and 1.344 eV as a function of excess energy for each emission energy. All three sets of fitting amplitude data show resonance features at 24 meV, and those at 1.344 eV were additionally enhanced at an excess energy of approximately 48 meV. The PL intensities are significantly weak under off-resonance conditions and increase with excitation energy above an excess energy of approximately 150 meV.

Here, we will discuss the origin of the resonance features observed in the PLE signal. One possible origin is the excited states ($n$=2, 3, …) of the exciton trapped in the moiré potential, where $n$ is the quantum number. We considered a simple model to calculate the energy levels of the moiré exciton in a parabolic potential trap with depth $U$, moiré period $a_{\text{moiré}}$=2 nm, and exciton effective mass $m$=0.25$m_0$, resulting in an interlevel spacing of $\Delta\hbar\omega = \hbar/a_{\text{moiré}}\sqrt{2U/m}$ in the harmonic-like potential[42]. In previous reports, the moiré potential depth $U$ was calculated to be approximately a few hundred meV in $R$-type stacking[19,25,43,44], and scanning tunnelling microscopy and spectroscopy (STM-STS) revealed a deep potential of a few hundred meV[45–47], which supports the observed onset of the PLE signal at approximately 150 meV in Fig. 2c and the PLE tail in Fig. 1a reflecting the continuum state of the moiré exciton ($n=\infty$). By assuming a moiré potential depth of $U$~200-300 meV, the interlevel spacing is estimated as $\Delta\hbar\omega$~160-200 meV, which is much larger than the observed energy differences of resonance peaks of 24 and 48 meV. In addition, by considering a zero-point energy $\Delta\hbar\omega/2$ in the parabolic potential, the excited states of the confined state ($n=2, 3, …$) are outside the potential, and only the ground state of the moiré exciton ($n=1$) is confined in the moiré potential. This estimation is consistent with the more sophisticated first-principles-based exciton moiré potential calculated with the plane-wave expansion method, which reports that the highly excited states ($n=2, 3, …$) are allowed only in heterobilayers with a very small twist angle of a few degrees[44]. Therefore, the excited states of the moiré exciton cannot explain the observed PLE resonances. The absence of the excited states of the moiré exciton suggests that the observed several peaks originate from the inhomogeneous energy distribution of different potential minima in the twisted heterobilayer.

The absence of the excited states of the moiré exciton is in contrast to the 2D intralayer exciton system, where the resonance signals of the excitonic Rydberg series of 2*s*, 3*s*, … are prominent in the PLE spectrum[48,49]. In the moiré system, the confinement by the in-plane moiré potential makes the exciton a 0D-like system, suggesting a drastic change from the Rydberg excitons in the non-hydrogenic 2D potential[25]. Indeed, as shown in the Supplementary Information, no prominent resonance features were observed at the estimated Rydberg energy in the non-hydrogenic 2D potential, which suggests that the moiré superlattice works as an in-plane potential and confines the interlayer exciton, causing a drastic change in the exciton energy level structure.

The energy difference of 24 meV (~200 cm$^{-1}$) observed in the PLE resonance is close to the energy of the A$_{1g}$ phonon mode at 30 meV (~240 cm$^{-1}$) in MoSe$_2$, WSe$_2$ and their heterobilayer observed in the Raman scattering spectrum[50], suggesting that phonon resonances are candidates to explain our result. This is similar to the frequently observed phonon resonances in the PLE spectra of 0D quantum dot systems[51,52]. Although the energy difference of 24 meV is slightly smaller than the energy of the A$_{1g}$ phonon mode, the difference might come from the following. The phonon modes around the Γ point are excited in the Raman scattering process because of the very small wavenumber of light. On the other hand, this is not the case with the interaction between the moiré exciton and phonon. Because the moiré excitons are localized in real space, the wavefunction of moiré excitons in momentum space is expected to be spread. In this situation, the phonon mode at any wavenumber as well as around the Γ point could interact with the moiré exciton, which implies that the energy of the interacting phonon is affected by the density of states of the phonon mode. Because the A$_{1g}$ phonon mode has the highest frequency at the Γ point, which becomes lower in the rest of momentum space from the dispersion curve, the averaged energy of the A$_{1g}$ mode would shift to the lower energy side with consideration of the phonon density of states[53], which is consistent with our measurement. In addition, because A$_{1g}$ phonon modes with various momenta contribute to the resonance process of photoexcitation in the PLE signals, the observed linewidth is expected to be broader than that in the Raman scattering spectrum, which is consistent with the observed relatively broad resonant peaks in Fig. 2c. This is in contrast to the intralayer exciton in the 2D system, where sharp phonon resonances are experimentally observed[29,54]. Then, we concluded that the experimentally observed resonant peaks at 24 and 48 meV are assigned to the interaction of the moiré excitons with one- and two-phonon modes, respectively.

The interactions with phonons are expected to cause characteristic transfer and relaxation processes of moiré excitons between the different moiré potential minima. To obtain information on the dynamic process, we measured the excitation power dependence of the PL spectra under an excitation of 1.55 eV, which is below the intralayer exciton energy, as shown in Fig. 3a. At low excitation intensities below ~3 kW/cm$^2$, we observed that the PL spectra show several emission peaks and that the spectral shapes do not change. Moreover, the whole spectral bandwidth is relatively small at ~25 meV. As the excitation power increases above approximately 12 kW/cm$^2$, PL signals emerge on the higher energy side at approximately 1.36 eV, and the centre of gravity shifts to the higher energy side, accompanied by broadening and feature-less behaviour. Figure 3b shows the PL intensity observed for several peaks (red, yellow, green, and blue marks in Fig. 3a) as a function of the power density. All peaks linearly increase with the excitation power under low excitation conditions below $10^4$ W/cm$^2$ and then follow saturation behaviour under higher excitation conditions. In addition, the lower energy peaks show saturation behaviour at lower excitation powers: The PL intensities at 1.321, 1.333, and 1.345 eV saturate at approximately $10^4$ W/cm$^2$, while that at 1.362 eV saturates at approximately $10^6$ W/cm$^2$.

Because only the ground state ($n$=1) is confined and the observed PL peaks originate from the energy level distribution of the ground states in the moiré potential, the PL spectral shapes under different excitation power densities provide information on the moiré exciton transfer and filling in the moiré potential. Under low excitation power conditions below 3 kW/cm$^2$, the invariant spectral shape from 1.32 to 1.345 eV with a spectral bandwidth of ~25 meV means that excitons occupy the energy levels within the bandwidth but do not occupy the potential minima with higher energy. This suggests that the relaxation and transfer of the moiré excitons from higher to lower energy levels with an energy separation $\Delta E$ below ~25 meV is not efficient and that the energy levels with small energy separation are decoupled, while the relaxation and transfer from higher to lower energy levels with an energy separation $\Delta E$ of ~25 meV is efficient. The schematics of possible relaxation processes are illustrated in Fig. 3c, d, e. Because the energy separation of $\Delta E$~25 meV is close to the phonon energy of $E_{phonon}$~24 meV, one-phonon emission is considered to play a dominant role in the relaxation due to the moiré exciton discrete density of states, as shown in Fig. 3c. When the energy difference of the initial and final potential minima in the relaxation process does not well correspond to the phonon energy, the relaxation is suppressed because of the absence of phonon scattering, as shown in Fig. 3d. On the other hand, under high excitation power conditions of approximately 12 kW/cm$^2$, the emergent higher energy components above the bandwidth in the PL spectra mean that the excitons gradually occupy the potential minima with higher energy. This suggests that relaxation with an energy separation of ~25 meV is also suppressed, implying state filling of the energy levels[55] because of the occupation of each moiré potential minimum with low energy by two optically generated excitons, as shown in Fig. 3e. At a saturation power density of 10$^4$ W/cm$^2$, the moiré potential minima with energy levels within a bandwidth of ~25 meV are almost filled with the photoexcited excitons at $5\times10^{10}$ cm$^{-2}$ (see Supplementary Information for the estimation of the exciton density).

This scenario is verified by considering the opposite case of coupled energy levels, where the fast relaxation process of the moiré exciton between different potential minima compared with the PL lifetime occurs regardless of the energy separation. This would result in different spectral shapes under various excitation power conditions: only one peak with the lowest energy is expected at a low excitation power because of the relaxation process to the potential minimum with the lowest energy, and as the excitation power increases, new peaks would appear beginning with the lowest energy levels because of the finite density of states and state filling effect, as shown in Fig. S4. This behaviour could not explain the experimental observation of several peaks with invariant spectral shapes at low excitation power in Fig. 3a, suggesting the existence of different relaxation processes of moiré excitons depending on the energy separation.

Based on the proposed relaxation process of the moiré exciton between different potential minima in Fig. 3c, we numerically simulated the PL spectra under different exciton densities. We assumed that the relaxation of the moiré exciton for energy separation equal to (different from) the phonon energy (~24 meV) is faster (slower) than the PL lifetime of ~10 ns and a Gaussian distribution of energy levels (see Supplementary Information for the PL lifetime). The relaxation process of the moiré exciton between different potential minima depends on the exciton occupation number of each potential minima considering the state filling effect due to Pauli blocking. We performed Monte Carlo simulations and simulated spectra using the linewidth extracted from the experimental data, as shown in Fig. 3f. The potential minimum density for different energies was estimated with the saturation power density, while the total density of moiré potential minima can be estimated to be ~$10^{13}$ cm$^{-2}$ based on the moiré period of 2 nm. The simulated spectra well reproduce the behaviour of the experimentally observed PL spectra in Fig. 3a: the nearly invariant spectral shape at low exciton density and the shift of the centre of gravity to the higher energy side at higher exciton density. Figure 3g shows the simulated intensity as a function of exciton density. The PL intensities at 1.320, 1.335, and 1.342 eV saturate around the exciton density of $10^{10}$-$10^{11}$ cm$^{-2}$, while that at 1.362 eV saturates around $10^{12}$ cm$^{-2}$. The good agreement between the experiment and the simulation suggests that the relaxation of the moiré exciton between the different potential minima is mainly caused by the efficient phonon emission process in the moiré potential.

In summary, we provide the dynamics of the moiré exciton and the moiré exciton and phonon interaction in a twisted WSe$_2$/MoSe$_2$ heterobilayer. The PLE spectra reveal that tuning the excitation energy can allow selective excitation of the moiré exciton of specific energy levels through the phonon interaction, although several peaks originate from the distribution of energy levels because of the inhomogeneity. The $\delta$-function discrete energy levels with only the exciton ground state of the moiré exciton in the heterobilayer with a relatively large twist angle are manifested in the negligibly small off-resonant PLE signal and the absence of excited states ($n$=2, 3, …). The moiré exciton dynamics between different potential minima with discrete energy levels were revealed to be assisted by the resonant phonon scattering process, and the relaxation via phonon scattering is suppressed to be slower than 10 ns at off-resonance with the phonon energy. Our results shed light on new aspects of moiré exciton and phonon coupling and lay the groundwork to explore quantum phenomena in moiré superlattices for quantum emitters with extremely low threshold lasing and so on.

This work was supported by JSPS KAKENHI (Grant No. JP16H00911, JP15K13337, JP15H05408, JP16H00910, JP16H06331, JP17H06786, JP17K19055, 19K14633, JP20H02605, and JP20H05664), JST CREST (Grant No. JPMJCR16F3 and JPMJCR18I5), the Keihanshin Consortium for Fostering the Next Generation of Global Leaders in Research (K-CONNEX) established by the Human Resource Development Program for Science and Technology, MEXT, and the Asahi Glass Foundation. Growth of $h$-BN was supported from the Elemental Strategy Initiative conducted by the MEXT, Japan, Grant No. JPMXP0112101001, JSPS KAKENHI Grant No. JP20H00354 and the CREST(JPMJCR15F3), JST.

**Figure captions**

**Figure 1. a** Schematic of the moiré potential and trapped exciton. **b** $\delta$-function discrete density of states of the moiré system. **c** Optical image of the 1L-$MoSe_2$/1L-$WSe_2$ heterobilayer. **d** PL spectrum of the $MoSe_2$/$WSe_2$ heterobilayer under excitation of 2.32 eV (red curve), and PLE spectrum monitored at ~1.35 eV (blue curve). **e** PL spectrum at excitation power densities of 1.3 kW/cm$^2$ (red curve) and 0.7 W/cm$^2$ (blue curve; intensity scaled by 60×) with a photon energy of 1.959 eV.

**Figure 2. a** PL spectrum obtained under near-resonant excitation from 1.356 to 1.377 eV measured at a power density of 13 kW/cm$^2$. The black lines are fitted with multiple Lorentz functions. **b** 2D PLE intensity plots. The tilted black dashed lines show excess energies of 24 and 48 meV. **c** PLE spectrum after Lorentzian fitting of the PL spectrum as a function of excess energy. The linewidth of the PLE signal is ~3 meV. The grey filled areas show excess energies of 24 and 48 meV.

**Figure 3. a** Normalized PL spectra for different excitation power densities under an excitation photon energy of 1.55 eV with an energy resolution of 3 meV. **b** PL intensity as a function of excitation power density at different probe energies. **c, d,** and **e** Schematics of the relaxation process of the ground state moiré exciton (*n*=1) between the different potential minima with energy separation $\Delta E$ equal to the phonon energy $E_{phonon}$, with $\Delta E$ different from the phonon energy, and under the state filling effect due to Pauli blocking at a high excitation density, respectively. **f** Simulated PL spectra for different exciton densities. **g** Simulated intensity as a function of exciton density.

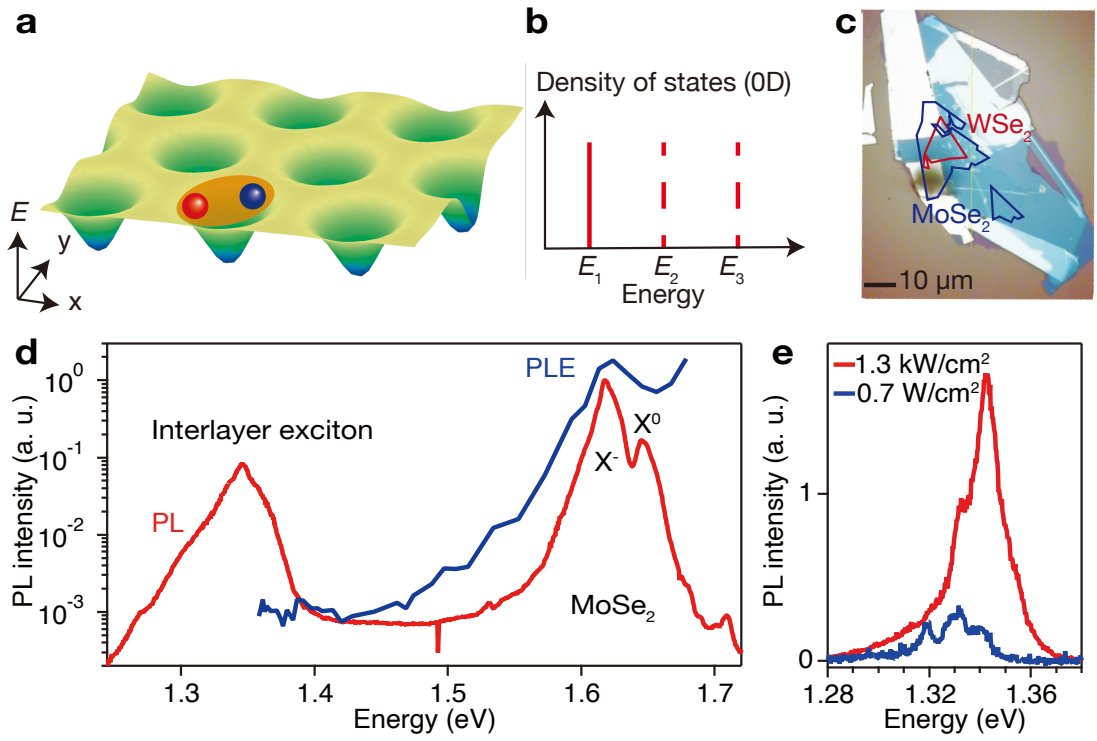

Figure 1 K. Shinokita *et al.*

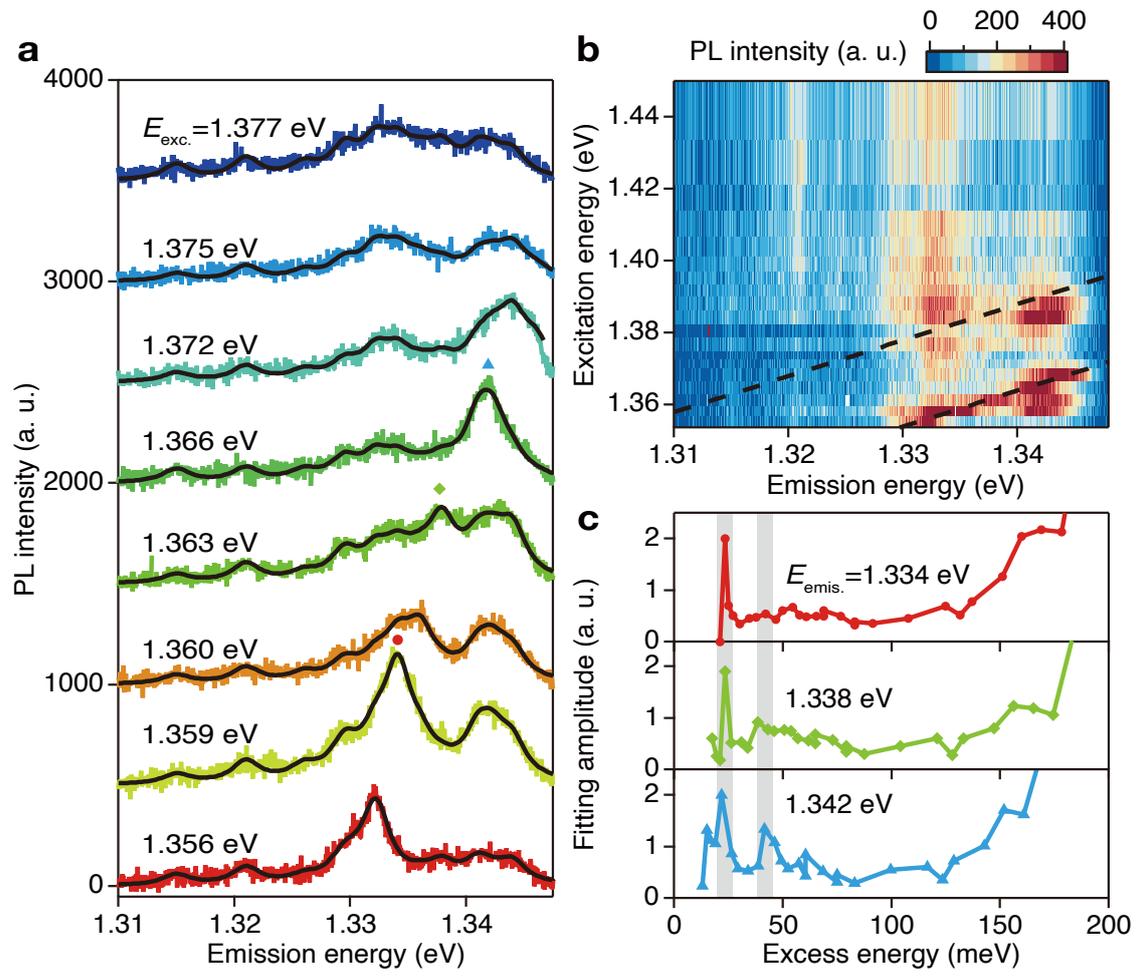

Figure 2 K. Shinokita *et al.*

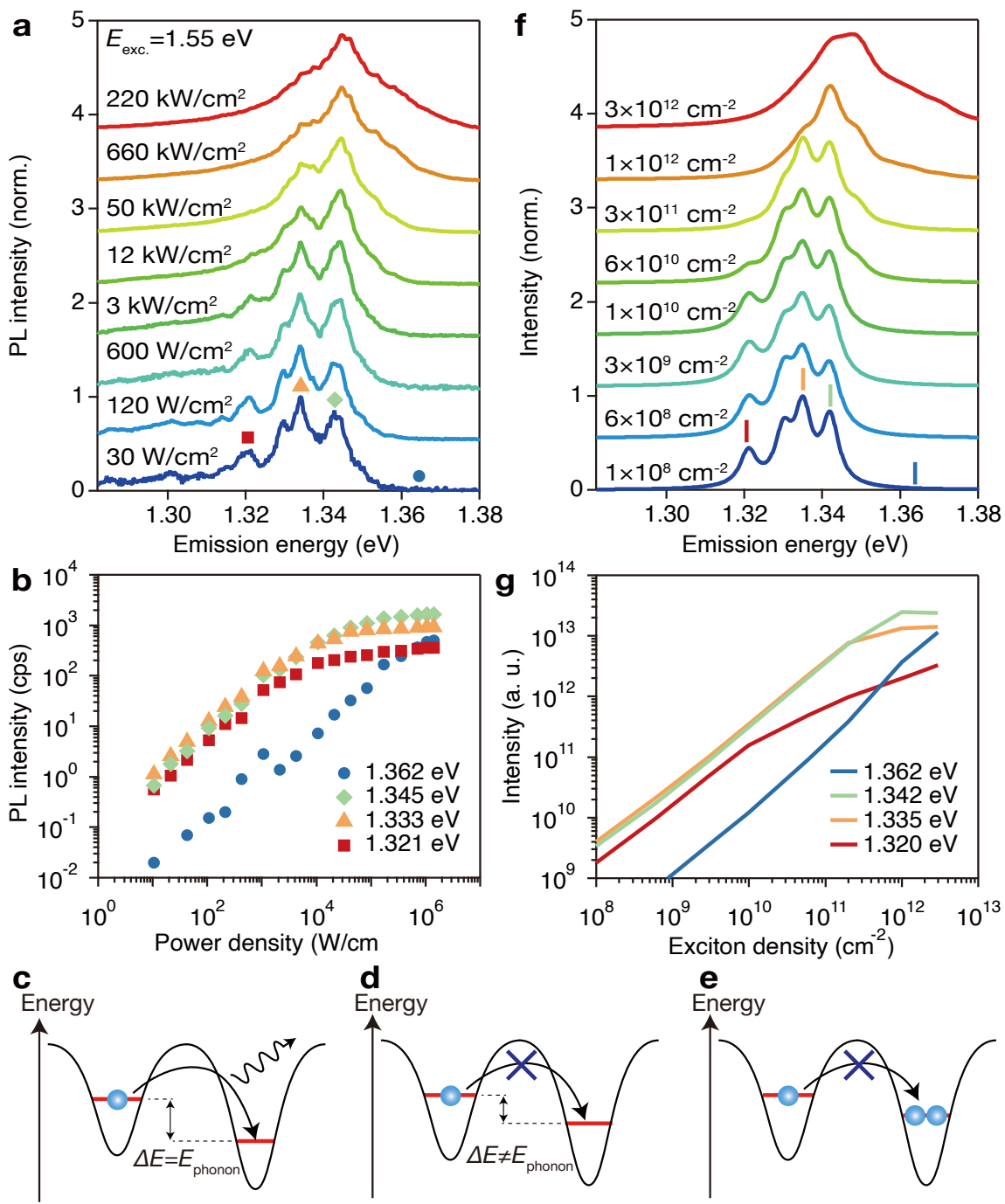

Figure 3 K. Shinokita *et al.*